\documentclass[fleqn,twoside]{article}
\usepackage{epsfig,espcrc2}

\usepackage{graphicx}
\usepackage[figuresright]{rotating}

\newcommand{\AmS}{{\protect\the\textfont2
  A\kern-.1667em\lower.5ex\hbox{M}\kern-.125emS}}

\newcommand{\text}[1]{\mbox{#1}}

\title{Auger : A large Air Shower Array and Neutrino Telescope}

\author{A. Letessier-Selvon\address[LPNHE]{Laboratoire de Physique Nucl\'eaire et des Hautes \'Energies, 
        Tour 33 Rez de Chauss\'ee, \\ 
        4, place Jussieu 75252 Paris Cedex 05, France}}%
       
\begin{document}

\begin{abstract}
Detection of Ultra High Energy Neutrinos (UHEN), with energy above 0.1 EeV~($10^{18}$ eV)
is one of the most exciting challenges
of high energy astrophysics and particle physics. In this article we show that the Auger Observatory, 
built to study ultra high energy cosmic rays, is one of the most sensitive neutrino telescopes 
that will be available 
during the next decade. Furthermore, we point out that the Waxman-Bahcall upper bound for high 
energy neutrino flux below 1 EeV turns into a lower bound above a few EeV.
In this framework and given the experimental evidences for $\nu_\mu\longrightarrow\nu_\tau$ with large mixing, 
we conclude that 
observation of Tau UHEN in the southern Auger observatory should most certainly occur within the next five years.
\vspace{1pc}
\end{abstract}

\maketitle

\section{Introduction}
%UHECR problematics, motivations for neutrino detection}
The nature and origin of the observed Ultra High Energy Cosmic Rays (UHECR) have been for the past decades the subject of numerous debates and models (for a review see e.g. \cite{SiglRev,BoratavRev,WatsonRev}). None of those models seemed to be able to 
fulfill simultaneously the source power requirement, its invisibility and the transport energy losses
without requiring either new physics or large magnetic field together with a local over-abundance of transient (power sufficient?) sources.
Moreover 
recent experimental results\cite{HiresICRC,AgasaICRC} shown at the 27th ICRC conference\cite{ICRC2001}, 
seemed somewhat contradictory in particular concerning the flux of events above the Greisen Zatsepin and 
Kuzmin cut-off\cite{GZK}. Given the lack of statistics the interpretation of the same data has even 
led some authors to completely contradictory conclusions (see e.g. \cite{WB0206217,BGG0204357}). 
What are the UHECR sources and their distribution ? What is their nature and energy spectra ? What is the flux 
above $10^{20}eV$ ? Those are the few fundamental questions that future experiments need to answer.
\par
In the energy range [$10^{19}-10^{20}$] eV only 
stable hadrons (e.g. protons) or nuclei, among the known particles, can travel on distance much larger than a few tens of Mpc. 
However the maximum distance is still limited to about 50 Mpc (our neighborhood on cosmological scale) above $10^{20}$ eV because of 
photo-production or photo-dissociation processes against the Cosmic Microwave Background of radiation (CMB). 
If astrophysical objects such as AGN or GRB are to play a fundamental role in UHECR production then particle spectra from distant
sources will be affected by these processes. Resulting fluxes should extinct exponentially 
above $10^{20}$ eV exhibiting the GZK cutoff. Taking into account additional distortions induced for example by extra 
galactic magnetic fields we conclude that hadron spectra in this energy range will mix the source characteristics 
 together with the transport distortions. 
On the other hand neutrinos can travel on cosmological distances essentially unaffected and all 
astrophysical UHECR hadron sources as well as more exotic ones such as topological defect collapses 
or heavy relic decays are bound to produce Ultra High Energy Neutrino (UHEN). Their detection would be a very valuable 
clue to the characteristics of the UHECR sources.

\section{Auger}
Because of their very low flux UHECR cannot be observed directly from space. 
Instead one must reconstruct the primary cosmic ray characteristics from the Extended Air Shower (EAS) it produces 
when interacting with the atmosphere.  
The Auger Observatories~\cite{AugerDesign} consist of two instrumented sites. One in the southern hemisphere,
now under construction in Malarg\"ue (Argentina), and the other one in the northern hemisphere. 
The detection system combines the two major techniques : a fluorescence detector system to measure the 
longitudinal profile of the EAS and a surface array of detectors to sample its lateral distribution at the ground. 
With a detection acceptance larger than 16,000~km$^2$sr per site Auger should observe each year more than 
5,000 events above $10^{19}$ eV, 500 above $5\times10^{19}$ and more than 100 above $10^{20}$.

\begin{figure}[t]\label{hybrid}
\epsfig{file=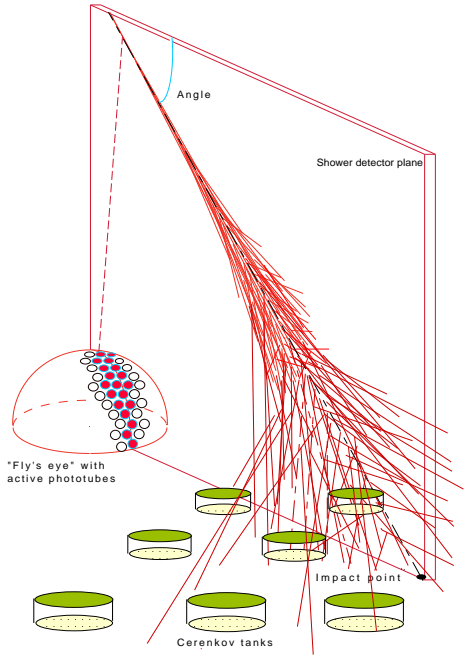, width=60mm}
\vspace*{-1cm}
\caption{Hybrid operation mode of the Auger detectors.}
\end{figure}

\begin{figure}[!t]
\begin{center}
\hspace*{-1.80cm}
\includegraphics[width=10cm]{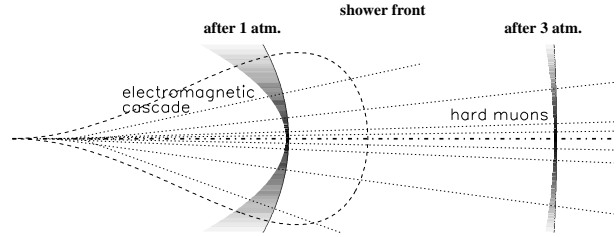}
\vspace*{-1.2cm}
\caption{Horizontal shower development.}
\label{showerdev}
\end{center}
\end{figure}

The southern Auger site will be composed of 1600 Cherenkov stations (our surface detector units) and 
4 fluorescence eyes located at the periphery of the array. Each eye is composed of six  $30^{o}\times30^{o}$ mirror and camera 
units looking inwards over the surface station network.
During year 2001 we accomplished all of our planned objectives : complete the construction of the 
assembly, central acquisition and office, and first fluorescence detector buildings; 
construct and deploy 40 Cherenkov stations (31 equipped with electronics) and develop and install two fluorescence 
mirror/camera prototype units. 
This mini Auger, called Engineering Array (EA) 
has been successfully running since then and has shown excellent performances most of them being above our 
initial specifications. Many events were recorded including about 100 hybrid events seen by both 
the fluorescence and the surface detectors (see Figure~1).  
It is however beyond the scope of this paper to report on these data and on the detector performances which will be the subject of a dedicated
publication\cite{Inprep}. Let us mention however, that the
construction of the observatory is going on well. A new fluorescence building is now completed and 
a pre-production observatory with 2 complete eyes and 140 stations should be running in 2003.

\begin{figure}[!t]
\begin{center}
\hspace*{-0.8cm}\includegraphics[width=9cm]{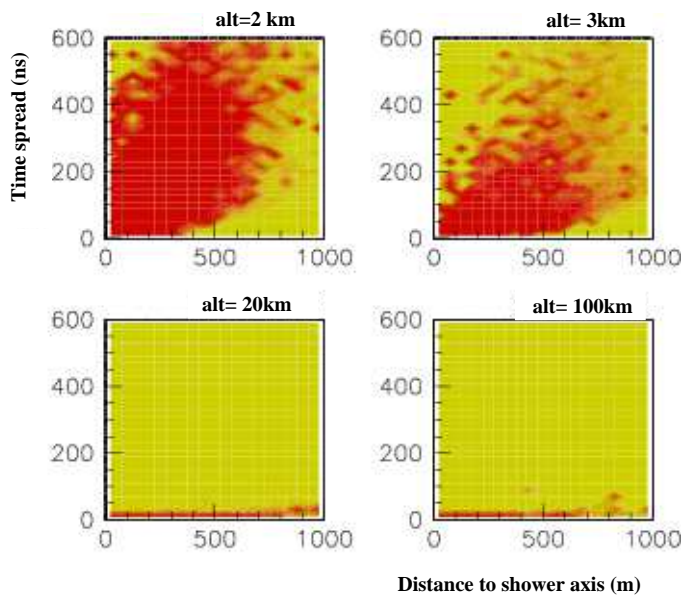}
\vspace*{-1cm}
\caption{Particle time spread with respect to a planar shower front versus distance to the shower axis for $10^{19}$eV protons at 80 deg. zenith angle. The primary altitude is given at the interaction point, early interactions (bottom)
correspond to high altitude and produce old shower at the ground level, late interactions (top) correspond to
penetrating particle and young shower. }
\label{discrim}
\end{center}
\end{figure}

\begin{figure}[!t]
\epsfig{file=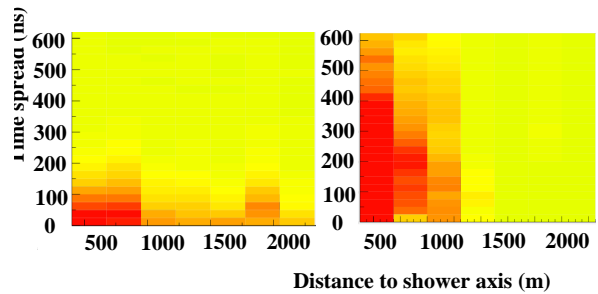,width=100mm}
\vspace*{-1cm}
\caption{Same as Figure~\ref{discrim} but for real data, the time resolution (vertical bins) is 25 ns in this case.
To the the left a most probably hadronic horizontal (therefore old) shower with a zenith angle of about 82 deg. To the right a near 
vertical (therefore young) shower with zenith angle of 45 deg.}
\label{discrimData}
\end{figure}

\section{Neutrino Identification and Acceptance}
Previous studies on UHEN interaction in the atmosphere and observation with Auger were reported 
in \cite{Zas2,Billoir}. UHEN may be detected and distinguished from ordinary hadrons by the
shape of the horizontal EAS they produce. Ordinary hadrons have large cross sections and interact very early
after entering the atmosphere. At large
zenith angles (above 80~deg.) the particles propagation distance along the shower axis 
between the shower maximum and the ground becomes larger
than 100~km. At ground level the electromagnetic part of the shower is totally extinguished 
(more than 6 equivalent vertical atmosphere were gone through) and only high energy muon survive
(Figure~\ref{showerdev}).
In addition, the shower front is very flat (radius larger than 100~km) and the particles time
spread is very narrow (less than 50~ns).
		      
\par
Unlike hadrons, neutrinos may interact deeply in the atmosphere and can initiate a shower in the volume 
of air immediately above the detector. These Horizontal Air Shower (HAS) will appear as ``normal young '' showers
with a curved front (radius of curvature of a few km), a large electromagnetic component, 
and with particles arrival well spread over time (over hundreds of nanoseconds) even at large distances from the axis.
These differences are striking when 
one looks at the ground particles time distribution versus the distance from shower axis. Figure~\ref{discrim} shows this distribution 
as computed 
from a Monte Carlo simulation while Figure~\ref{discrimData} compares two real events with widely 
different zenith angle as were recorded by the Auger EA. These plots should only be taken as indicatives. In particular 
 the horizontal shower in Figure~\ref{discrimData} has an estimated energy about five times larger 
than the vertical one, therefore it extends at much larger distance from the shower axis. A fair comparison would 
require a vertical shower of approximately the same energy whose larger lateral extension would allow to observe the difference 
in time spread over a wider range. Moreover the time bins in the real data is 25ns wide (this is not
a limitation since our tanks impulse response function is about 50 ns wide) while the Monte Carlo results were plotted 
with a 1 ns resolution. Nevertheless the difference is still quite clear.
\par
The acceptance to neutrino interactions in the atmosphere above the Auger array has been calculated in~\cite{Billoir,Guerard} 
and reaches about 40 km$^3$ water equivalent above a neutrino energy of $10^{19}$ eV. 
Although quite large this is not sufficient to see a few neutrino events per year except for some 
very speculative neutrino sources~\cite{Billoir,Guerard,Protheroe}. 
Even at these energies the neutrino interaction length in air is too large compared to the Auger array dimensions and 
our ``target'' does not convert enough neutrino into visible HAS above the detector.      

\section{Tau Neutrino detection}
The possibility to detect tau neutrinos in the Auger observatory was first introduced in\cite{Puebla2000} 
and was described in detail in a dedicated paper\cite{Bertou2002}. 
Unlike electrons which do not escape from
the rocks\footnote{If one does not take into account the LPM effect which significantly increases the electron path length above $10^{18}$~eV.} or muons that produce a much to faint signal in the atmosphere, taus, produced in the mountains or
in the ground around the Auger array, can escape even from deep inside the rock and produce a clear
signal if they decay above the detector.
\par
\begin{figure}[!t]
\begin{center}
\includegraphics[width=8.0cm]{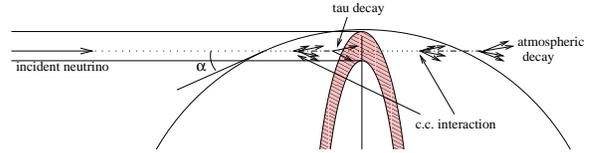}
\end{center}
\vspace*{-0.50cm}
\caption{Chain of interactions producing an observable tau shower.}
%\vspace*{0.5cm}
\label{simul}
\end{figure}

The chain of reactions together with the geometrical 
configuration that must be met to produce a detectable tau shower are depicted in Figure~\ref{simul}.
These conditions are rather severe : the neutrino must be almost perfectly
horizontal (within less than  5 deg.) the tau produced by its interaction should escape the earth and then decay over a 
distance matching the Auger surface array  dimension, finally the decay must occur at low altitude. 
These criteria can be met as a result of a number of favorable numerical coincidences :
\begin{itemize}
\item At $10^{18}$~eV the tau decay length is of the order of the Auger array dimension (50 km).
\item In earth, the neutrino interaction length at this same energy is a few 100 km.
\item For skimming angle below 5$^o$ neutrino propagate through a few 100~km of earth a distance corresponding to their 
      interaction length 
\item A few $10^{18}$~eV is the neutrino energy corresponding to the threshold of pion photo-production by protons on the CMB.
\end{itemize}
A Monte-Carlo technique has been used to simulate the tau neutrino or charged lepton
interactions and propagation inside the Earth.
The lepton may interact several times through deep inelastic scattering, changing charge in most cases,
or eventually decay, but, in all cases, a tau neutrino or charged lepton is
present in the final state. Some energy is lost at each interaction,
as well as continuously along the paths. However, in our energy range, the
initial direction of the incoming neutrino is always
conserved (Figure~\ref{simul}).

We computed that most of the detectable signal (90\%)
comes from upward going $\nu_\tau$ where the interactions
occur in the ground all around
the array and only 10\% from downward going $\nu_\tau$ coming from
interactions in the mountains surrounding the array.

\begin{figure}[!t]
\hspace*{-0.8cm}\includegraphics[width=8.5cm]{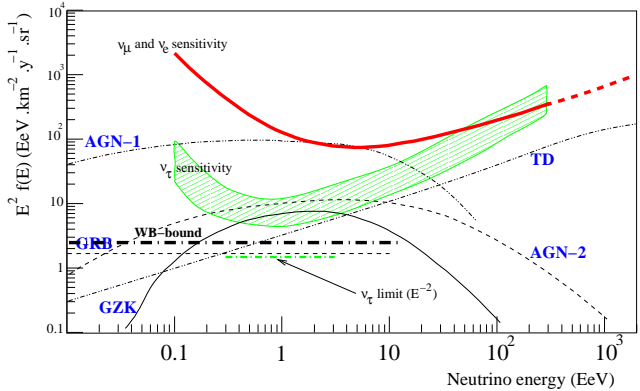}
\vspace*{-1.3cm}
\caption{Muon or tau neutrino and anti-neutrino fluxes
from various sources in the full mixing hypothesis, taken
from~\cite{Protheroe} (and divided by 2). Dotted lines are
speculative fluxes, dashed probable and solid certain.
The thick solid line and the hatched area represent the Auger sensitivity
defined by $I_{10}(E)=1$, i.e. one event per year and per decade.
Top line for horizontal shower from $\nu_e$ and $\nu_\mu$ interactions in the atmosphere
hatched area for tau induced showers under strong DIS loss (top of the area) or no DIS loss (bottom).
Any flux lying above those curves for at least one decade will give
more than one events per year in Auger. The WB-bound is well within our sensitivity.
We also plotted the 90\% C.L. limit (background free detection) for an E$^{-2}$ flux between 0.3 and 3 EeV
that Auger could achieve after five years.}
\label{taufig}
\end{figure}

The results that we obtained 
are summarized in Figure~\ref{taufig} and in Table~\ref{table} and are very encouraging, even in the case of the highest contribution of 
Deep Inelastic Scattering (DIS-high) to the tau interaction length (reducing the length over which the tau can escape from inside the earth).
In case of a null result (although unexpected!) Auger could 
set after five years a limit as low as  $5\times10^{-9}$~GeV\,cm$^{-2}$s$^{-1}$sr$^{-1}$ at 90\% CL 
assuming an $E^{-2}$ energy dependent flux in the range [0.3,3] EeV.

\section{The WB upper bound as a lower limit}
In 1998, Waxman and Bahcall have calculated a neutrino flux upper limit from astrophysical transparent source, now 
referred to as the WB limit (or WB bound)~\cite{WBbound}. Some authors~\cite{WB-controversy} have argued that this limit was too restrictive 
in particular because of the spectral properties of the sources.
In this section we do not wish to enter this debate but rather follow 
Waxman and Bahcall's argument to conclude that even for those sources a detectable neutrino signal should 
be visible in Auger.  

\par
In their paper the authors suppose that UHECR are protons produced by astrophysical sources. The source flux 
is then normalized to the observed cosmic ray flux assuming that those protons  do not loose most of their energy 
at the production point against the local photon background.  This is the transparency condition which is deduced 
from the fact that those sources are visible in gamma rays. At this stage a neutrino flux upper limit
is obtained assuming that all the proton energy gets converted into neutrino. 

\par
Below the pion photo-production threshold against the CMB background (around $7\times10^{19}$ eV) 
the computed neutrino fluxes are, by construction, upper limits for those sources. 
However, for protons with energy above the  threshold the situation is quite different~: they will 
effectively convert their energy into neutrino on distances of the order of a few tens of Mpc. 
Therefore, if those sources are the dominant contribution to the observed cosmic rays in the range [10,100]~EeV and 
since most of them should lie outside our GZK sphere they will produce a neutrino flux of the order of the WB bound.
\par
Since it is generally admitted that any other non transparent astrophysical source would directly produce more neutrinos and
since in all top down models the neutrino flux at the production point is much larger then the proton flux, we can conclude 
that for neutrino energy above a few\footnote{The exact location in energy depends on the source distribution dependence to 
red-shift.} $10^{18}$~eV the neutrino flux should be above the WB bound. 

\par
The WB bound, as given in reference~\cite{WBbound}, is
$1.5\times10^{-8}$~GeV\,cm$^{-2}$s$^{-1}$sr$^{-1}$ for a strong z-evolution of the sources. 
Only half of those neutrino may convert into $\nu_\tau$ on their way to Earth given a flux of 
 $7.5\times10^{-9}$~GeV\,cm$^{-2}$s$^{-1}$sr$^{-1}$. This flux is fifty percent larger than the Auger $\nu_\tau$ null result limit and
 within our sensitivity (Figure~\ref{taufig}). An example of cosmogenic neutrinos produced by protons from astrophysical 
 source (taken from~\cite{Protheroe}) is shown under the GZK label in Figure~\ref{taufig} with the corresponding number of events in Table~\ref{table}.

\begin{table}[!t]
\caption{Expected number of events after five years for the source models presented
in Fig.~\ref{taufig} and various DIS contributions to continuous energy losses.}\vspace*{0.2cm}
\begin{center}
\begin{tabular}{l|c|c|c|c|c}
DIS & AGN-1 & TD & GRB & GZK & AGN-2 \\ \hline
% BS & 38.4 & 2.1 & 0.7 & 2.4 & 3.9
none & 135 & 11.5 & 2.5 & 8.5 & 14.5 \\ \hline
low  & 120  & 9.0 & 2.0 & 7.5 & 12.5 \\ \hline
high & 50.0 & 4.0  & 1.0 & 3.0 & 5.5
\end{tabular}\vspace*{-0.2cm}
\end{center}
\label{table}
\end{table}

\section{Conclusions}
The Auger observatory is a very sensitive neutrino telescope reaching for a null result a limit in $\nu_\tau$ flux as low as 
$5\times10^{-9}$~GeV\,cm$^{-2}$s$^{-1}$sr$^{-1}$ at 90\% CL. Moreover, given the observation of UHECR in 
the range [10,100] EeV, given the lower limit on neutrino flux which can be derived from the WB bound and  
given the strong experimental evidence for \mbox{$\nu_\mu\longleftrightarrow\nu_\tau$} with large mixing the expectations for a positive 
result in the next five years are very high. Beside its excellent performances for the study of UHECR Auger may very well be  
the first experiment to observe  $\nu_\tau$ appearance from a $\nu_\mu$/$\nu_e$ source.

\end{document}